\documentclass[useAMS,usenatbib]{mn2e}

\usepackage{graphicx}
\usepackage{times}
\title[Cameron band emission in comets]{Model for Cameron band 
emission in comets: A case for EPOXI mission target comet 
103P/Hartley 2  }
\author[Anil Bhardwaj and Susarla Raghuram]{Anil Bhardwaj \thanks{E-mail:
anil\_bhardwaj@vssc.gov.in; bhardwaj\_spl@yahoo.com} and Susarla Raghuram
\thanks{E-mail: raghuramsusarla@gmail.com}
\\  Space Physics Laboratory, Vikram Sarabhai Space Center, Trivandrum, 695022, India\\
 }
\begin{document}

\date{ }

\pagerange{\pageref{firstpage}--\pageref{lastpage}} \pubyear{2002}

\maketitle

\label{firstpage}

\begin{abstract}
The CO$_2$ production rate has been derived in comets using the Cameron band 
(a$^3\Pi$ - X$^1\Sigma$) emission of CO molecule assuming that 
photodissociative excitation of CO$_2$ is the main 
production mechanism of CO in a$^3\Pi$ metastable state. 
We have devoloped a model for the production and loss of CO(a$^3\Pi$) which has
been applied to comet 103P/Hartley 2: the target of EPOXI mission. Our model calculations
 show that photoelectron impact excitation
 of CO and dissociative excitation of CO$_2$ can together contribute about 60-90$\%$ to the 
Cameron band emission. The modeled brightness of (0-0) Cameron band emission on comet
 Hartley 2 is consistent with Hubble Space Telescope  observations 
for 3-5\% CO$_2$ (depending on model input solar flux) and 0.5\% CO relative to water, where 
photoelectron impact contribution
is about 50-75\%.  We suggest that estimation of CO$_2$ abundances on comets using Cameron 
band emission may be reconsidered. We predict the height integrated column brightness of Cameron 
band of $~\sim$1300 R during EPOXI mission encounter period.
 
\end{abstract}


\section{Introduction}
\label{sect1}
In the exploration  of the solar system, comets have been targeted by 
various space missions. After successful encounter of comet 9P/Temple 1 on 4 July  2005, the NASA's 
Deep impact mission, also called EPOXI mission, under it's extended investigation program DIXI will 
encounter comet 103P/Hartley 2 on 4 November 2010, with closest approach around 700 km from the nucleus. 
This comet has been observed by several space telescopes in different spectral regions 
\citep{Weaver94,Crovisier99,Colangeli99,Groussin04,Lisse09,Snodgrass08,Snodgrass10}. 

The first clear detection of the Cameron band (a$^3\Pi$ - X$^1\Sigma$) of CO was
reported by \cite{Weaver94} in HST/FOS spectrum of comet 103 P/Hartley 2.
Since Cameron band emission is a forbidden transition, resonance fluorescence is 
not an effective excitation mechanism. The upper state of this emission (a$^3\Pi$) is 
a metastable state with lifetime of 3 ms \citep{Gilijamse07}, which is quite small.
Thus, Cameron band emissions are treated as ``prompt emissions'' and can be used to 
probe distribution of 
parent species if this emission is produced in dissociative excitation of a molecular
 species. Photodissociative excitation of CO$_2$ is considered as the major production 
mechanism of CO Cameron band and has been used to trace the distribution and abundance of 
CO$_2$ on comets \citep{Feldman97,Weaver94}.

Besides photodissociative excitation of CO$_2$, there are other channels of excitation of 
  CO molecule in the  a$^3\Pi$ state. It has been shown that photoelectrons generated
by solar EUV radiation also play an important role in excitation, dissociation, and ionization
  processes leading to emission and chemistry in cometary comae 
\citep[e.g.,][]{Ip86,Boice86,Korosmezey87,Bhardwaj90,Bhardwaj96,Bhardwaj99a,Bhardwaj03,Weaver94,
Haider05,Capria08}. 
Recently, \cite{Campbell09} demonstrated the importance of photoelectron impact excitation   
in comets, and showed that electron impact on CO gives 40\% contribution to the total CO 
Fourth positive emission. The presence of photoelectron excitation in cometary coma is clearly publicised 
by detection of OI 1356 \AA \ emission in comets \citep[e.g.,][]{Sahnow93,Mcphate99}, since this emission 
being a spin-forbidden transition cannot be produced by solar fluorescence. There are other 
siginificant evidences 
for an important role of photoelectron excitation in cometary coma
 \citep[e.g.,][]{Tozzi98,Bhardwaj99a,Feldman09}.
 In addition to photon and electron impact reactions, dissociative electron recombination reactions of 
CO$^+$-bearing ions can also produce CO in the a$^3\Pi$ excited state.
 
Our aim in this paper is to study various production and loss mechanisms of CO(a$^3\Pi$)
and to estimate the contribution of  photoelectron impact excitation of CO and CO$_2$ in the
production of Cameron band for different relative abundances of CO$_2$ on comet
103 P/Hartley 2: the target of EPOXI mission. Since model calculations  depend on 
 input solar flux, we have  estimated its sensitivity on 
the calculated intensity of  Cameron band emission. We show that photoelectron impact on CO and 
CO$_2$ are dominant 
processes ($\sim$60--90\% contribution) in producing CO molecule in (a$^3\Pi$) state. Around the 
EPOXI encounter epoch
predictions are made for the brightness of Cameron band on for comet 
103P/Hartley 2, which will be
observed by several space-based telescopes including HST.
\section {Model}
\label{sect2}
We have developed a model for the production of Cameron band emission on comets, which uses the basic coupled 
chemistry model described in detail in our earlier papers \citep{Bhardwaj96, Bhardwaj99a, Haider05}.
Various sources of production and loss of CO(a$^3\Pi$) 
are summarized in Table \ref{prodlosstab}.
The total water production rate is taken as 6.3 $\times$ 10$^{28}$ s$^{-1}$ for 
comet 103P/Hartley 2 \citep{Weaver94}.  

 To evaluate the effect of solar EUV flux on model 
calculations, we have considered 2 solar flux models: EUVAC model of \cite{Richards94} and  SOLAR 2000 (S2K) 
model of \cite{Tobiska00}.  The 
degradation of the solar UV-EUV radiation and solar EUV-generated photoelectrons in 
the coma is modeled using the method of \cite{Bhardwaj90,Bhardwaj96} and further 
developed by \cite{Bhardwaj99a,Bhardwaj03}. 
The electron impact production rates are calculated using the Analytical Yield 
Spectrum (AYS) approach, which is based on the Monte Carlo method. Details of AYS 
approach are given in several of the previous  papers \citep{Bhardwaj90,Bhardwaj96, 
Bhardwaj99a,Bhardwaj99d, Bhardwaj99b,Bhardwaj09}. The current model 
takes into account the most recently published cross 
sections for the photon impact and  electron impact dissociation, ionization, 
and excitation processes for the gases in the coma. The cross section for 
photodissociative excitation of CO$_2$ producing CO in a$^3\Pi$ state is calculated 
using absorption cross sections of CO$_2$ and  yield of Cameron band measured by \cite{Lawrence72}.
 The cross section for electron impact excitation  of 
CO(a$^3\Pi$) from CO is taken from \cite{Jackman77} and for dissociative 
excitation of CO$_2$  is taken from \cite{Bhardwaj09}. 
The electron temperature profile 
required for dissociative recombination reactions is taken from \cite{Korosmezey87} and is 
assumed to be same as on comet Halley. 
Calculations are made for the comet 103P/Hartley 2 at heliocentric
distance of 0.96 AU. 
\section{Results and Discussion}
\label{sect3}
The photodissociation of CO$_2$ producing CO in a$^3\Pi$ state is determined by solar flux mainly
in the wavelength region 550 to 1050 \AA. 
Table~\ref{a3piprodtab}  presents the calculated photon production frequencies of CO(a$^3\Pi$)  for two 
different solar flux 
models. The CO(a$^3\Pi$) production frequencies calculated for photoelectron impact on CO$_2$ and CO are also 
shown in the same table for the corresponding   solar flux models.  Our calculated photodissociation 
frequencies 
are about 50\% to a factor of 2 lower than those reported by \cite{Huebner92}.

\begin{figure}   
\centering 

 \noindent\includegraphics[width=18pc,angle=0]{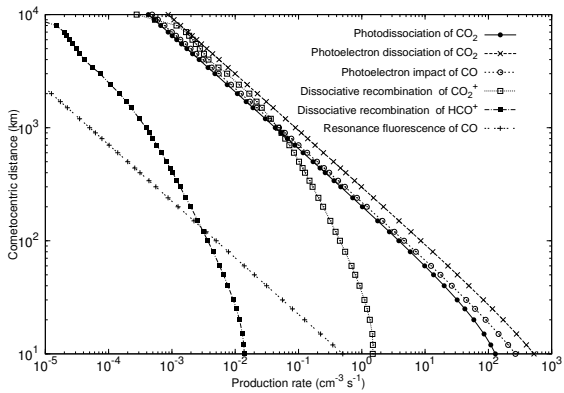}   
\caption[figure1]{Radial profiles of the production processes of 
CO(a$^3\Pi$) in comet 103P/Hartley 2 for 0.5\% CO and 4\% CO$_2$ using EUVAC solar flux on 18-19  sept 1991.}
\label{proda3pi}
\end{figure}

\begin{figure}    
\centering 
 \noindent\includegraphics[width=18pc,angle=0]{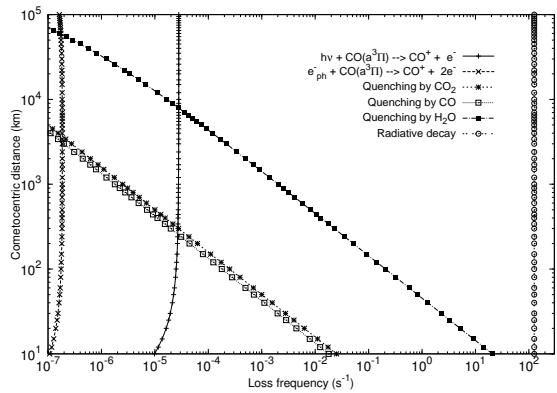}  
\caption {Radial profiles of the loss frequencies of CO(a$^3\Pi$) 
in comet 103P/Hartley 2 for 0.5\% CO and 4\% CO$_2$ using EUVAC solar flux on 18-19 sept  1991.}
\label{lossa3pi}
\end{figure}

Using EUVAC solar  flux, the calculated radial profile of volume production rate 
for the various sources of CO(a$^3\Pi$) at the relative abundance of 4\% CO$_2$ and 
0.5\% CO are shown in Figure~\ref{proda3pi}. At 100 km cometocentric distance, the dominant source of
 production of CO(a$^3\Pi$) is electron 
impact of CO$_2$ ($\sim$50\%) followed by electron impact of CO ($\sim$25\%), and 
photodissociation of CO$_2$ ($\sim$15\%). The contributions 
from dissociative recombination reactions are quite small ($\le$5\%) at lower
 cometocentric distances, but the recombination of CO$_2^+$  is a  significant ($<$30\%) source
at 1000 km and beyond. Figure~\ref{lossa3pi} shows radial profile of various loss processes of CO(a$^3\Pi$) 
 for the same relative composition of CO$_2$ and CO. Since lifetime of CO in excited state 
(a$^3\Pi$) is very short \citep[$\sim$3 ms;][]{Gilijamse07}, the radiative decay is the 
dominant loss process. Collisional quenching of CO(a$^3\Pi$) by cometary neutral
species is negligible since 103P/Hartley 2 is a low production rate comet. But in the case 
of large production rate comets, like 
Hale-Bopp,  quenching by water would be a dominant loss process in the innermost part of the coma.

  Figure~\ref{proja3pi} shows the modeled limb brightness profiles of Cameron band emission for 
different production processes of CO(a$^3\Pi$). The cometary coma is assumed to be 
spherically symmetric. The production rates are integrated 
up to 10$^5$ km along the line of sight at a given projected distances from the cometary 
nucleus, and converted into brightness. The brightness profiles are averaged over the projected 
area of slit (2870 $\times$ 954 km) corresponding to the HST observation \citep{Weaver94}. 
The volume emission rate for 3 transitions (0-0, 1-0, 0-1) of Cameron band emission 
are calculated using the following formula
\begin{equation}
V_{\nu'\nu''}(r)=q_{o\nu'} (A_{\nu'\nu''}/\sum_{\nu''}A_{\nu'\nu''})\ V(r)\ exp(-\tau)
\end{equation}
where V(r) is total volume excitation rate of CO(a$^3\Pi$) at cometocentric distance r,
q$_{o\nu'}$ is the Franck-Condon factor 
for transition, A$_{\nu'\nu''}$ is Einstein transition probability from upper 
state $\nu'$  to lower state $\nu''$, and $\tau$ is optical depth. Since resonance 
fluorescence is not an effective excitation mechanism for the Cameron band and the total
gas production rate is only 6.3 $\times$ 10$^{28}$ s$^{-1}$,
the cometary coma can be safely  assumed to be optically thin. The Franck-Condon 
factors are taken from \cite{Nicholls62} and branching ratios from \cite{Conway81}.
 The relative contributions of
(1-0), (0-0), (0-1) transitions to the total Cameron band are 13.9\%, 10.4\%, and 14.7\%, respectively.

Table~\ref{bigtab} presents the model calculated slit-averaged brightness of (1-0), (0-0), (0-1)
transitions of Cameron band, as well as total Cameron band brightness and height-integrated 
column brightness for different relative abundances of CO and CO$_2$ 
 corresponding to the HST observation of comet 103P/Hartley 2 on September 18-19, 1991. 
Due to the absence of CO Fourth positive emission  in this comet \citep{Weaver94}, the abundance
 of CO is constrained to 0.5\%. However, we do consider a case of 1\% of CO to evaluate it's implications on 
the results. 
This table also depicts fractional contribution of photodissociation of CO$_2$, 
photoelectron impact of CO and CO$_2$, and dissociative recombination of 
CO$_2^+$ to the total calculated brightness at 3 projected distances (10$^2$, 
10$^3$, and 10$^4$ km) from the nucleus.
Since the production rates of photodissociative excitation of CO$_2$, and photoelectron 
impact of CO and CO$_2$, are dependent on input solar flux model, results are presented for 
the EUVAC and S2K solar  fluxes relevant to the date of comet observation which was in solar
maximum condition. 

 The HST observation of 0-0 transition of Cameron band is 35 Rayleigh \citep{Weaver94}, which is consistent 
with 
model calculated brightness for the relative abundance of 4 to 5\% of CO$_2$ and 0.5\% CO when EUVAC solar
flux is used. In this case, at 100 km, the photoelectron impact of CO$_2$ (50\%) and CO (25\%) contribute
 around 75\%, while photodissociative excitation of CO$_2$ is $<$15\%. At 1000 km and beyond,  
the contribution due to electron impact of CO$_2$ and CO is about 60--70\% while those   
of dissociative recombination of 
CO$_2^+$ is $\sim$15--30\% and  of photodissociative
 excitation of CO$_2$ $\sim$10\% only.  On an average, the  photoelectron impact of CO$_2$ and CO  
 contribute  about 60-75\% to the  production of Cameron band emission, while 
photodissociative excitation of CO$_2$ contribute about 10-15\% only. 
 In the case of S2K solar flux  model, the CO$_2$ abundance of 3 to 4\% is required to match HST-observed 
Cameron band 0-0 transition brightness. Here  the contribution of photodissociative excitation
 of CO$_2$ is  $\sim$20\%, while the electron impact of CO$_2$ and CO together  contribute 
  $\sim$65\%, to the total Cameron band emission. When the CO abundance is doubled to 1\% of water 
the relative contribution due to electron impact on CO increases resulting in the reduction (by $\sim$1\%) 
in the requirement for CO$_2$ abundace to match the HST-observation brightness. However, there is no major
 change in the  percentage contributions due to photodissociation and photoelectron impact 
excitation of CO and CO$_2$.
Hence, we conclude that the photodissociative 
excitation of CO$_2$ is not the dominant source for the production 
of Cameron band in comets.

\section{Summary}
    The summery of Table~\ref{bigtab}  and the above results is that whatever be the relative 
abundances of CO$_2$ and CO, and the solar input flux, the photodissociation contribution is in the
 range of 5--20\%, while
photoelectron impact excitation contribution is about 50--80\%. We conclude that 
photoelectron impact excitation of Cameron band in comets is much more important than
was thought previously. Further, the contribution due to photoelectron impact excitation of CO 
is significant and vary between 20-50\%. We have also shown that dissociative recombination of 
CO$_2^+$ ion is also 
a significant source of Cameron band at cometocentric distances $\ge$1000 km  with contribution of 
as high as 30\%.  Hence, the derivation of CO$_2$ abundances
based on the assumption that photodissociation of CO$_2$ is major mechanism should be
revisited. The results presented in this paper suggest that Cameron band emission may not be used
to probe cometary CO$_2$ abundance uniquely.
\section{Prediction}
\begin{figure}   
\centering 
 \noindent\includegraphics[width=18pc,angle=0]{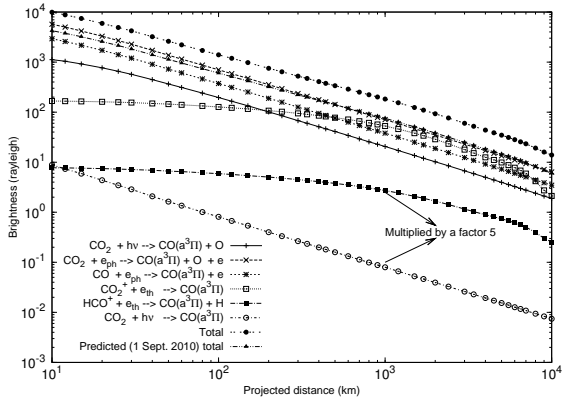}
\caption {Brightness profile of Cameron band on comet 103P/Hartley 2 as a function
 of projected distance from nucleus for 4\% CO$_2$ and 0.5\% of CO using EUVAC solar flux on 
18-19 sept 1991. Resonance 
fluorescence values and dissociative recombination profiles
 are plotted after multiplication by a factor 5. Total
brightness predicted profile for EPOXI mission period is also shown}
\label{proja3pi}
\end{figure}

The EPOXI mission will encounter the comet 103P/Hartley 2 on 4 November 2010 when comet will be at 
heliocentric distance 
of 1.06 AU and geocentric distance 0.15 AU. In the current solar minimum period, using our model 
we have estimated 
the intensity of Cameron band  around mission 
encounter date by taking solar flux on 1 sept. 2010 for both EUVAC and S2K solar flux models. 
The calculated Cameron band intensity is tabulated in table~\ref{bigtab}  for the relative abundance of 
4\% of CO$_2$ and 0.5\% of CO. For EUVAC solar flux, 
the predicted height-integrated column intensity of Cameron band is 1365 R, in which 
1-0, 0-0, and 0-1 
 band transitions contribute 189, 142, 200 R, respectively. Figure~\ref{proja3pi} presents  
the total Cameron band predicted brightness profile as a function of  projected distances from the nucleus.
For the same relative abundances of CO$_2$ and CO and  the solar S2K flux, the predicted intensity is 1258 R.
The results presented in the paper will help in 
understanding and interpretation of the extensive data to be obtained during EPOXI encounter period.

\renewcommand{\thefootnote}{\fnsymbol{footnote}}

  \begin{center}
\begin{table*} 
\caption{Reactions for the production and loss of CO($a^3\Pi$)}
\label{prodlosstab}
\begin{tabular}{|llp{1.7in}|}
\hline
\multicolumn{1}{|c}{ Reaction}
&\multicolumn{1}{l}{ Rate(cm$^{-3}$ s$^{-1}$ or s$^{-1}$)}
&\multicolumn{1}{c|}{ Reference} \\
\hline
CO$_2$ + h$\nu$ $\rightarrow$ CO(a$^3\Pi$) + O($^3$P)
&Model & \cite{Schunk09, Lawrence72} \\
CO + h$\nu$ $\rightarrow$ CO(a$^3\Pi$)
&1.69$\times$ 10$^{-9}$ &\cite{Weaver94}\\
CO$_2$ + e$^-_{ph}$ $\rightarrow$ CO(a$^3\Pi$) + O + e$^-$
&Model&\textit{Present work}\\
CO + e$^-_{ph}$  $\rightarrow$ CO(a$^3\Pi$) + e$^-$ 
 &Model &\textit{Present work}\\
CO$_2^+$  + e$^-$ $\rightarrow$ CO(a$^3\Pi$) + O  
&{6.5 $\times$ 10$^{-7}$ (300/Te)$^{0.8}$ $\times$ 0.87\footnotemark[2]
 $\times$ 0.29\footnotemark[3]} &
\cite{Seiersen03,Rosati03},\\
HCO$^+$ + e$^-$ $\rightarrow$ CO(a$^3\Pi$) + H
&0.23\footnotemark[4] $\times$ 2.4 $\times$ 10$^{-7}$ (300/Te)$^{0.7}$
&\cite{Rosati07, Schmidt88}\\
CO(a$^3\Pi$) + h$\nu$ $\rightarrow$ C + O
&7.2$\times$ 10$^{-5}$  & \cite{Huebner92} \\
CO(a$^3\Pi$) + h$\nu$ $\rightarrow$ CO$^+$ + e$^-$
&8.58$\times$ 10$^{-6}$  & \cite{Huebner92} \\
CO(a$^3\Pi$) + h$\nu$ $\rightarrow$ O+C$^+$ + e$^-$
&2.45$\times$ 10$^{-8}$  & \cite{Huebner92} \\
CO(a$^3\Pi$) + h$\nu$ $\rightarrow$ C + O$^+$ + e$^-$
&2.06$\times$ 10$^{-8}$  & \cite{Huebner92}\\
CO(a$^3\Pi$) +H$_2$O $\rightarrow$ CO + H$_2$O  &3.3 $\times$ 10$^{-10}$  
& \cite{Wysong00}\\
CO(a$^3\Pi$) + CO$_2$ $\rightarrow$ CO + CO$_2$  &1.0 $\times$ 10$^{-11}$  
&\cite{Skrzypkowski98} \\
CO(a$^3\Pi$) + CO $\rightarrow$ CO + CO  &5.7 $\times$ 10$^{-11}$ 
&\cite{Wysong00}\\
CO(a$^3\Pi$) + e$^{-}_{ph}$ $\rightarrow$ CO$^{+}$ + 2e$^{-}$  & Model
&\textit{Present work}\\
 CO(a$^3\Pi$)\hspace{0.3cm}  $\longrightarrow$ CO  +  h$\nu$ & 1.6 $\times$ 10$^{-9}$  
& \cite{Weaver94} \\
\hline \\
\end{tabular}
\addtocounter{footnote}{1}

\footnotemark{\small{0.87 is yield of dissociative recombination 
of CO$_2^+$ producing CO}}; \footnotemark{\small{0.29 is yield of CO(a$^3\Pi$) produced from CO}};\\
 \footnotemark{\small{0.23 is yield of dissociative recombination 
of HCO$^+$ producing CO(a$^3\Pi$); e$^-_{ph}$ = photoelectron.}}
\end{table*}
 \end{center}

\begin{center}

 \begin{table*}

 \caption{Production frequency (s$^{-1}$) of 
CO(a$^3\Pi$) for three different processes  at 1 AU}
\begin{minipage}{150mm}
  \begin{tabular}{|c|c|c|c|c|c|c|c|c|c|c|c|c|c|c|c|c|c|}
\hline
\label{a3piprodtab}
Solar flux&\multicolumn{2}{{p {1.5 in}|}}{Photodissociation of CO$_2$ }& \multicolumn{2}{{p {1.5 in}|}}
{Photoelectron impact of CO$_2$} &\multicolumn{2}{{p {1.5 in}|}}
{ Photoelectron impact of CO}\\
 \cline{2-3}  \cline{4-5} \cline{6-7}
&Solar min & Solar max  & Solar min& Solar max & Solar min & Solar max \\ 
    &(2010 Sep 1)   &(1991 Sep 18)    & (2010 Sep 1)  & (1991 Sep 18) & (2010 Sep 1) &  (1991 Sep 18)  \\ 
\hline
S2K  &1.1  $\times$ 10$^{-7}$ & 5.6 $\times$ 10$^{-7}$ & 3.4 $\times$ 10$^{-7}$& 9.7 $\times$ 10$^{-7}$&
 1.5 $\times$ 10$^{-6}$& 4.3 $\times$ 10$^{-6}$ \\
EUVAC  & 1.7 $\times$ 10$^{-7}$ & 2.6  $\times$ 10$^{-7}$  & 3.5  $\times$ 10$^{-7}$  &8.9 $\times$ 10$^{-7}$
 & 1.5  $\times$ 10$^{-6}$  & 3.7 $\times$ 10$^{-6}$  \\

\hline
  \end{tabular}
\end{minipage}
 \end{table*}
\end{center}


\begin{center}
\begin{table*}   

\begin{minipage}{250mm}
\small
\caption{{ Calculated brightness of  Cameron band on comet 
103P/Hartley 2 for different conditions}}
\label{bigtab}
 \begin{tabular}{|c|c|c|c|c|c|c|c|c|c|c|c|c|c|c|c|c|c|c|c|c|}
\hline
\multicolumn{2}{|p{1 in}|}{\centering Relative abundance}
& \multicolumn{3}{p{0.9 in}|} {\centering \hspace{3cm}HST-slit averaged }    %
&\multicolumn{12}{p{3.5 in}|}{\centering {Percentage contribution to 
total Cameron band for different process at three 
different projected radial distances (km)}} 
&\multicolumn{2}{p{0.4 in}|}{\centering  Total Cameron band brightness (R)}
\\[-3pt]
\cline{1-2}
\cline{6-19} 
 \multicolumn{1}{|p{0.4 in}|} { \centering CO$_2$ (\%)} 
&\multicolumn{1}{p{0.4 in}|}{ \centering CO (\%)} 
&\multicolumn{3}{p{1.0 in}| }{  brightness (R)}  
&\multicolumn{3}{p{0.8 in}|} {\centering {Photodissociation of CO$_2$}  }  
&\multicolumn{3}{p{0.8 in}|} {\centering {Photoelectron impact of CO$_2$} }    
&\multicolumn{3}{p{0.8 in}|} {\centering {Photoelectron impact of CO}  }   
&\multicolumn{3}{p{0.8 in}|} {\centering {Recombination of CO$_2^+$} }
&\multicolumn{1}{p{0.1 in}|}{\centering HST-Slit Avg.}  & 
\multicolumn{1}{p{0.1 in}|}{\centering Height integrated nadir view}\\ [0pt]
\hline
 &  &\multicolumn{1}{p{0.25 in}|}{(1-0)} &\multicolumn{1}{p{0.25 in}|}{(0-0)} 
&\multicolumn{1}{p{0.25 in}|}{(0-1)}
&\multicolumn{1}{p{0.1 in}|}{10$^2$ }
&\multicolumn{1}{p{0.1 in}|}{10$^3$ }
&\multicolumn{1}{p{0.1 in}|}{10$^4$ }
&\multicolumn{1}{p{0.1 in}|}{10$^2$ } 
&\multicolumn{1}{p{0.1 in}|} {10$^3$ }
&\multicolumn{1}{p{0.1 in}|} {10$^4$ }
&\multicolumn{1}{p{0.1 in}|} {10$^2$ } 
&\multicolumn{1}{p{0.1 in}|} {10$^3$ }
&\multicolumn{1}{p{0.1 in}|} {10$^4$ }
&\multicolumn{1}{p{0.1 in}|} {10$^2$ } 
&\multicolumn{1}{p{0.1 in}|} {10$^3$ }
&\multicolumn{1}{p{0.1 in}|} {10$^4$ }
& & \\  
 \cline{3-17}
  \multicolumn{2}{|p{0.8 in}|}{ \centering EUVAC  \footnotemark[1]  } 
 &   &   &  &   &  &   &   &   &  &   &   &    &  &   &  && \\
 7  & 0.5 & 68  & 51 & 72 & 15  & 12 & 15  & 57  & 42 &  51  & 17  & 12  & 15  & 10& 32  & 17 & 489 & 4769\\
 5  & 0.5 & 51  & 39 & 54 & 14  & 11 & 14  & 53  & 40  & 48  & 22  & 17  & 21  & 9 & 30  & 16 & 369 & 3710\\
 4  & 0.5 & 43  & 32 & 45 & 13  & 11 & 13  & 50  & 38  & 45  & 26  & 20  & 25  & 9 & 29  & 15 & 309 & 3172\\ 
 3  & 0.5 & 34  & 26 & 36 & 12  & 10 & 12  & 46  & 36  & 42  & 32  & 26  & 30  & 8 & 26  & 13 & 250 & 2628\\
 2  & 0.5 & 26  & 20 & 28 & 11  & 9  & 10  & 39  & 32  & 36  & 42  & 34  & 40  & 7 & 23  & 12 & 190 & 2078\\
 4  & 1   & 51  & 38 & 54 & 11  & 9  & 11  & 40  & 32  & 36  & 41  & 33  & 38  & 7 & 24  & 12 & 311 & 3948\\
3  & 1   & 43  & 32 & 46 & 10  & 8  & 10  & 35  & 29  & 32  & 48  & 40  & 45  & 6 & 21  & 11 & 311 & 3420\\
  \multicolumn{2}{|p{0.9 in}|}{\centering S2K   \footnotemark[1] } 
   &   &   &  &   &  &   &   &   &  &   &   &    &  &   &  && \\[-3pt]
  7  & 0.5 & 86 & 64 & 91 & 26 & 21 & 25 & 48 & 36 & 44 & 15 & 11 & 14 & 10 & 31 & 16 & 622 & 5928\\
  5  & 0.5 & 65 & 48 & 68 & 25 & 20 & 24 & 45 & 35 & 42 & 20 & 15 & 19 & 9  & 29 & 15 & 467 & 4586\\
  4  & 0.5 & 54 & 40 & 57 & 23 & 19 & 22 & 43 & 33 & 40 & 24 & 19 & 22 & 8  & 28 & 14 & 390 & 3903\\ 
  3  & 0.5 & 43 & 32 & 46 & 22 & 18 & 21 & 40 & 31 & 37 & 30 & 23 & 28 & 8  & 26 & 13 & 313 & 3210\\
  2  & 0.5 & 33 & 25 & 35 & 19 & 16 & 18 & 35 & 28 & 32 & 39 & 32 & 37 & 8  & 23 & 11 & 236 &2508\\
  3  & 1   & 53 & 40 & 56 & 17 & 15 & 17 & 31 & 25 & 29 & 45 & 37 & 42 & 6  & 21 & 10 & 384 & 4093\\
  2  & 1   & 43 & 32 & 45 & 14 & 12 & 14 & 25 & 21 & 23 & 55 & 47 & 52 & 5  & 18 & 8 & 308 & 3411\\
  \multicolumn{2}{|p{0.8 in}|}{\centering Prediction   } 
&   &   &  &   &  &   &   &   &  &   &   &    &  &   &  &   &   \\
 \multicolumn{2}{|p{1 in}|}{\centering EUVAC \footnotemark[2]} 
   &  & &  &   &  &   &   &   &  &   &   &    &  &   &  &   &   \\
4  & 0.5 & 189\footnotemark[3] & 142\footnotemark[3] & 200\footnotemark[3] &  21 & 18 &  20 & 46 
 & 37  & 42 &  25 &  20 & 23  
 & 7 &  23 &  12& - & 1365  \\
 \multicolumn{2}{|p{0.8 in}|}{\centering S2K  \footnotemark[2]} 
  & & &  &   &  &   &   &   &  &   &   &    &  &   &  &  &   \\
 4  & 0.5 & 175\footnotemark[3] & 130\footnotemark[3] & 185\footnotemark[3] & 16  & 14 & 16  & 50  
& 41  & 46 & 28  &  23 & 
 25  & 7 & 23 & 13 & - & 1258  \\

\hline
\end{tabular}
\addtocounter{footnote}{-4}

\footnotemark{\small{Solar flux for 18 sept. 1991};}
\footnotemark{\small{Solar flux for 1 sept. 2010}}
\footnotemark{\small{Relative contribution of band in total nadir view intensity.}}
\end{minipage}
\end{table*}
 \end{center}


\label{lastpage}

\end{document}